\newcommand{\apjj}[2]{ApJ #1, #2}
\newcommand{\aeta}[2]{A\&A #1, #2}

\newcommand{\mn}[2]{MNRAS #1, #2}

\newcommand{\ergs}{erg s$^{-1}$}

\newcommand{\degree}{\degr}

\newcommand{\rx}{\object{RX\,J0720.4-3125}}
\newcommand{\rxa}{\object{RX\,J1856.5-3754}}

\documentclass{aa}
\usepackage{graphicx}
\usepackage{epsfig}
\begin{document}
   \title{The proper motion and energy distribution of the isolated neutron star \rx}

      \author{
       C. Motch 
      \inst{1}
      \fnmsep
      \thanks{Based on observations collected at the European Southern
Observatory, La Silla and Paranal, Chile (ESO Programmes 58.D-0596, 62.H-0706,
64.H-0156(A), 66.D-0286(A) and 70.D-0515(A)} 
      \and
      V.E. Zavlin
      \inst{2}
      \and
      F. Haberl
      \inst{2}
       }
   
   \offprints{C. Motch}

   \institute{
              Observatoire Astronomique, UA 1280 CNRS, 11 rue de l'Universit\'e,
              F-67000 Strasbourg, France
              \and 
              Max-Planck-Institut f\"ur extraterrestrische Physik, D-85740,
              Garching bei M\"unchen, Germany 
	      }
	      
   \date{Received ; accepted}

      \abstract{ESO 4m class telescope and VLT deep imaging of the isolated neutron star \rx \
reveals a proper motion of  $\mu = 97\pm12$\,mas/yr  and a blue U-B color index.  We show that a
neutron star atmosphere model modified to account for a limited amount of hydrogen on the star's
surface can well represent both the optical and X-ray data without invoking any additional
components. The large proper motion almost completely excludes the possibility that accretion from interstellar
medium is the powering mechanism of the X-ray emission. It also implies that the proposed spin
down  is entirely due to magnetic dipole losses. \rx \ is thus a very likely middle aged cooling neutron
star. Its overall properties are quite similar to some of the long period radio pulsars recently
discovered, giving further support to the idea that \rx \ may be a pulsar whose narrow  radio beam
does not cross the Earth.}

      %\keywords{tobedone }

   \maketitle
%
%________________________________________________________________

\section{Introduction}

ROSAT has identified a small group of hot and radio-quiet isolated neutron stars (INSs, see reviews by
Motch \cite{mo2001}, Treves et al. \cite{treves2000} and Haberl \cite{haberl03}). None of these objects
seems associated with a SNR and low interstellar absorption towards them indicates distances of the
order of a few hundred parsecs at most. Grating X-ray spectra of the two brightest members, \rx\
(Paerels et al. \cite{paerels2001}) and \rxa\ (Burwitz et al. \cite{bur2001}; Drake et al.
\cite{drake02}), do not significantly depart from pure blackbody energy distributions. Inferred
blackbody temperatures range between 50 and 100 eV. Soft X-ray pulsations with periods around 10\,s
are now detected in a majority of sources although some period candidates still require confirmation.
Optical counterparts are very faint objects whose brightness seems to be mostly due to thermal emission
from the neutron star surface (see Pavlov et al. \cite{pavlov2002} for a recent review). These INSs
undergo little, if any, magnetospheric activity and are thus prime targets to measure fundamental
parameters through modeling the neutron star thermal emission and to constrain the debated equations of
state. 

The evolutionary status of these objects is still unclear. Some of the 10$^{8}$ to 10$^{9}$ INSs born
in the Galaxy could have space velocities and rotation periods slow enough to accrete from relatively
dense phases of the interstellar medium and be re-heated to temperatures similar to those observed (see
e.g. Blaes \& Madau \cite{bm93}). The fraction of old INSs detectable in X-ray is of the order of
several percents but strongly depends on the assumed velocity distribution (Madau \& Blaes
\cite{madau94}). This hypothesis currently faces two main difficulties. First, the space density of
hot radio-quiet INSs is much lower than predicted by the accretion models (Treves et al.
\cite{treves2000}). Second, the rotation periods, although relatively long for neutron star standards,
are still too short to allow accretion to take place in average interstellar medium conditions unless
the magnetic fields of INSs have significantly decreased by about two orders of magnitudes over their
life time (see e.g. Wang \cite{wang97}). 

Alternatively, these INSs could be young cooling objects. Long rotation periods are observed in
few radio pulsars as well as in AXPs showing that under certain circumstances (efficient braking or
birth with a low angular momentum), relatively young neutron stars can rotate slowly.

Optical observations can efficiently help to discriminate between the two hypotheses, re-heating
by accretion from the insterstellar medium or relatively young cooling objects. For some of the 
objects, proper motions can be measured which, assuming Bondi-Hoyle accretion, may rule out the
accreting scenario. Additionally, the optical to X-ray flux ratio is sensitive to the chemical
composition of the neutron star surface (Pavlov et al. \cite{pavlov96}) and may therefore reveal the
presence of newly accreted material from the interstellar medium or from a fossil disc.

In both the optical and X-ray band, \rx\ is the second brightest INS among those discovered with ROSAT.
This object exhibits pulsations at a period of 8.39\,s (Haberl et al. \cite{haberl97}) with 
$|\dot{\rm P}|$ $<$ 3.6 10$^{-13}$\,s\,s$^{-1}$ (Kaplan et al. \cite{kaplan02}). The X-ray spectrum is
blackbody-like with kT$^\infty\simeq 86$ eV (as given by the ROSAT data). ESO-NTT and Keck observations
have led to the optical identification of \rx \ with a faint stellar-like object  (Motch \& Haberl
\cite{mh98}; Kulkarni \& van Kerkwijk \cite{KvK98}).We report here on new 4m-class telescope and ESO-VLT
observations  of \rx\ which reveal its proper motion, show its blue UV continuum and provide constraints
on the nature of its thermal emission. While this paper was in the refereeing process we became aware
of a preprint by Kaplan et al.(\cite{kaplan03}) reporting on HST observations of \rx \ which confirm the
steep optical to UV energy distribution of the star).

\section{Optical observations}

Optical imaging was obtained over five ESO periods spanning the time interval  from February 1997 till
January 2003. The log of observations is  given in Table~\ref{obslog}. Data collected in 1997 have
been already presented and discussed in Motch \& Haberl (\cite{mh98}). In January 1999 we used the
ESO-NTT and the SUSI2 imager equipped with EEV44-80 CCDs \#45 and \#46 (2048x4096 pixels). The
January 2000 run was carried out at the ESO 3.6m using EFOSC2 and the LORAL 2048$^{2}$ CCD \#40. At
UT1 we used in both cases FORS1 and a TK 2048$^{2}$ CCD. Individual exposures of $\sim$ 10\,mn were
obtained in 1997, 1999, 2000 and 2002-2003 for the B and R bands while the U band 2000 images were
15\,mn long. In January 1999 and 2000, the U and B filters were alternated and in all instances the
telescope was moved by a few arcseconds in different directions between two exposures. Raw images
were corrected for bias using over-scan regions or mean bias frames and corrected for flat-field
using dithered sky images obtained at dawn. Individual corrected images were then moved to a common
frame using a set of several nonsaturated bright stars and then stacked together applying a
statistical criterion in order to reject cosmic-ray impacts. Because of the excellent seeing
prevailing during ESO-VLT observations, the geometrical distortion effects of the FORS camera (up
to 0.2\arcsec \ on the edge of the field) combined with dithering slightly deteriorated the quality
of the stacked images. Distortion effects were thus corrected before stacking using the midas
rectify image task and parameters taken from the FORS users Manual (Issue 1.5). After correction,
the positional errors of reference stars dropped from 30\,mas to 5\,mas and the FWHM image size
sharpened by 0.1 to 0.05\arcsec.

\begin{table*}
\caption[]{Optical data}
\label{obslog}
\begin{tabular}{llccccc}
\noalign{\smallskip}
Date   &  Instrument & Band & Exp.  & Pixel  & seeing  \\
       & Telescope   &        & (ksec)&  (\arcsec )  & (\arcsec ) \\
\hline
 7-11 Feb 1997 &   SUSI/NTT &  U      & 7.2      &  0.13\arcsec & \\    
 7-11 Feb 1997 &   SUSI/NTT &  B      & 5.4      &  0.13\arcsec & \\
10-11 Mar 1997 &   SUSI/NTT &  U      & 3.6      &  0.13\arcsec & \\
   29 Mar 1997 &   SUSI/NTT &  B      & 1.8      &  0.13\arcsec & \\
   29 Mar 1997 &   SUSI/NTT &  U      & 1.8      &  0.13\arcsec & \\
    4 Apr 1997 &   SUSI/NTT &  B      & 1.8      &  0.13\arcsec & \\
\hline    
    Sum B 1997 &   SUSI/NTT &  B      & 9.0      &  0.13\arcsec & 0.91\arcsec\\
    Sum U 1997 &   SUSI/NTT &  U      & 12.6     &  0.13\arcsec & 0.97\arcsec\\
\hline
   15 Jan 1999 &  SUSI2/NTT &  B      & 12.0     &  0.157\arcsec & 0.80\arcsec\\ 
   15 Jan 1999 &  SUSI2/NTT &  R      & 6.0      &  0.157\arcsec & 0.82\arcsec\\ 
\hline
   7-8 Jan 2000 & EFOSC2/ESO-3.6m &  B      & 17.4     &  0.314\arcsec & 0.94\arcsec\\
   7-8 Jan 2000 & EFOSC2/ESO-3.6m &  U      & 26.1     &  0.314\arcsec & 0.94\arcsec\\
\hline
20-23 Dec 2000 &  FORS1/VLT-UT1 &  B      & 24.1     &  0.10\arcsec  & 0.60\arcsec\\ 
\hline 
28 Dec 2002 - 2 Jan 2003 &  FORS1/VLT-UT1 &  B  & 24.8  & 0.10\arcsec  & 0.5 - 1.2\arcsec\\  
\noalign{\smallskip}
\hline
\end{tabular}
\end{table*}
     
\section{Optical photometry}

\begin{figure}
\psfig{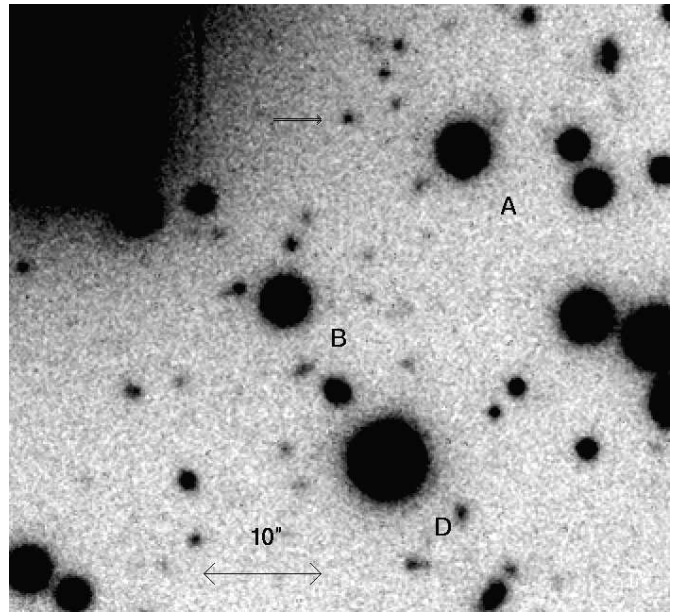}

\caption[]{Sum of best seeing B band images obtained with FORS1 on UT1 in December 2002 and January
2003. Positions of the photometric reference stars are shown. North is to the top and East is to
the left. The arrow marks the position of \rx }

\label{fc}
\end{figure}

\begin{figure}
\psfig{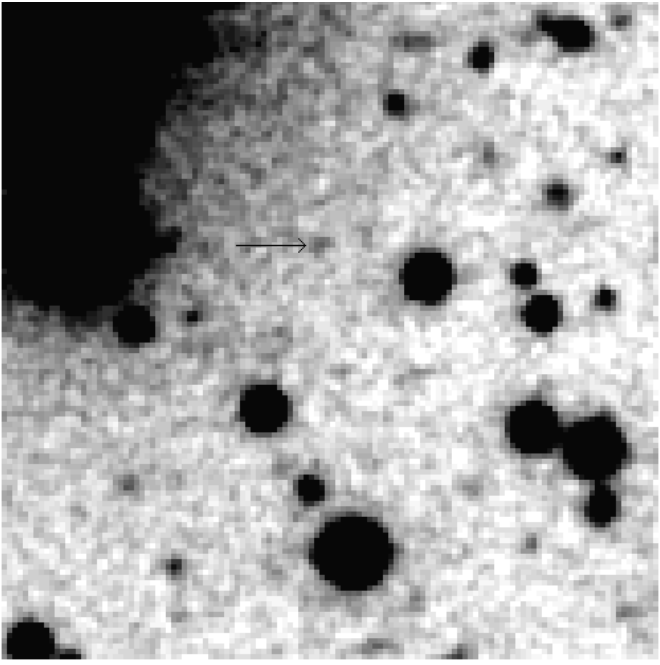}

\caption[]{The  summed U band image obtained with EFOSC2 on the ESO-3.6m in January 2000. The area
covered is roughly the same as that seen in Fig.~1. The arrow shows the position of \rx}

\label{fcu}
\end{figure}

Fig. \ref{fc} shows the deep B image obtained by summing the 3.4\,h FORS1  observation in 2002-2003 of best
image quality resulting in an average FWHM seeing of 0.56\arcsec. The optical counterpart of \rx\ (object "X1"
from Motch \& Haberl (\cite{mh98}) or "X" in Kulkarni \& van Kerkwijk (\cite{KvK98})) is clearly detected,
thanks to the excellent images and VLT throughput. Our relatively deep U exposure obtained by summing the
7.2\,h long observation in January 2000 collected at the ESO-3.6m (see Fig. \ref{fcu}) confirms the strong
UV excess of the counterpart seen by Kaplan et al. (\cite{kaplan03}). Object "X2" in Motch \& Haberl
(\cite{mh98}) does not appear exactly at the same position in the 1997 image as in the Keck observation of
Kulkarni \& van Kerkwijk (\cite{KvK98}) or as in our 1999, 2000 January and VLT images. We believe that a low
amplitude unfiltered sky background fluctuation could be the origin of its displacement. The revised position
of "X2" outside of the ROSAT error box and its lower UV flux excludes it as a possible counterpart of \rx\
(Kulkarni \& van Kerkwijk \cite{KvK98}). Our R image in January 1999 is not deep enough to detect the
counterpart.  

Absolute photometric calibration was obtained in the course of the January 2000 run. Standard stars
in the fields of PG0231+051, PG0942-029 and PG1047+003 (Landolt \cite{landolt}) were used to derive U
and B magnitudes of three comparison stars named A, B and D located close to \rx . Stars A and B are the
same as in Haberl et al. (\cite{haberl97}). Their magnitudes are listed in Table \ref{UBstd} and their
positions are shown in Fig. \ref{fc}. The sky background close to \rx \ is somewhat contaminated by stray
light from a nearby group of relatively bright stars. Because of its equatorial mounting images obtained
at the ESO 3.6m further suffer from fixed diffraction spikes  caused by the secondary spider. Following 
Kulkarni \& van Kerkwijk (\cite{KvK98}), we found that the most accurate photometric measurements were
achieved by fitting a 2-D Gaussian distribution of fixed width on  top of an inclined plane representing
the sky background and stray light from nearby stars. The fixed width of the Gaussian profile was derived
from bright  nonsaturated field stars and the photometric scale measured on averaged images since the
flux from bright stars is not accurately conserved in the image stacking process. Colour transformations
provided by ESO were systematically applied in order to compensate for the large colour differences
between \rx \ and the photometric field reference stars. Magnitudes derived from the various runs are
listed in Table \ref{magrx}. The 1999-2003 B magnitude of \rx \ is consistent with that reported by 
Kulkarni \& van Kerkwijk (\cite{KvK98}), B = 26.6 $\pm$ 0.2, and the re-analysis of the 1997 images using
2-D Gaussian fitting instead of aperture photometry also provides  consistent values, B = 26.58 $\pm$
0.25. The U-B index is very blue, U-B = -1.11 $\pm$ 0.18, assuming no B optical flux variation.  The
simultaneous but less accurate B measurement in January 2000 yields a consistent U-B = -0.76 $\pm$ 0.23.
Our estimate of the U band flux agrees well with the optical to far UV distribution measured by Kaplan et al.
(\cite{kaplan03}). The difference of 0.17 magnitude in B between the two ESO-VLT observations is only a
2.6 $\sigma$ effect and is thus  probably not significant. We also used the high quality optical data
from the last ESO-VLT run to search for night to night variability. There again, we find some scatter
with a maximum amplitude of 0.25 mag from one night to the other which is only significant to a 2.9
$\sigma$ level. It is more likely that the rather large photometric scatter is due to a slight
underestimate of the actual photometric errors. 

\begin{table}
\caption[]{Magnitudes of field reference stars (1$\sigma$ errors)}
\label{UBstd}
\begin{tabular}{ccc}
\noalign{\smallskip}
Star   &  U  & B  \\ 
\hline
A      &  20.585 $\pm$ 0.014 & 20.542 $\pm$ 0.009 \\
B      &  21.011 $\pm$ 0.015 & 20.841 $\pm$ 0.009 \\
D      &  18.770 $\pm$ 0.014 & 18.766 $\pm$ 0.008 \\
\noalign{\smallskip}
\hline
\end{tabular}
\end{table}

\begin{table}
\caption[]{U and B magnitudes of \rx \ (1$\sigma$ errors)}
\label{magrx}
\begin{tabular}{ccc}
\noalign{\smallskip}
Epoch          &   U              & B \\ 
\hline
February-April 1997 &              & 26.58 $\pm$ 0.25 \\
January  1999   &                  & 26.79 $\pm$ 0.20 \\
January  2000   & 25.68 $\pm$ 0.17 & 26.44 $\pm$ 0.15 \\
December 2000   &                  & 26.787 $\pm$ 0.040 \\
Dec 2002 - Jan 2003   &                  & 26.620 $\pm$ 0.050 \\
\noalign{\smallskip}
\hline
\end{tabular}
\end{table}

\section{Proper motion study}

\begin{figure}

\psfig{figure=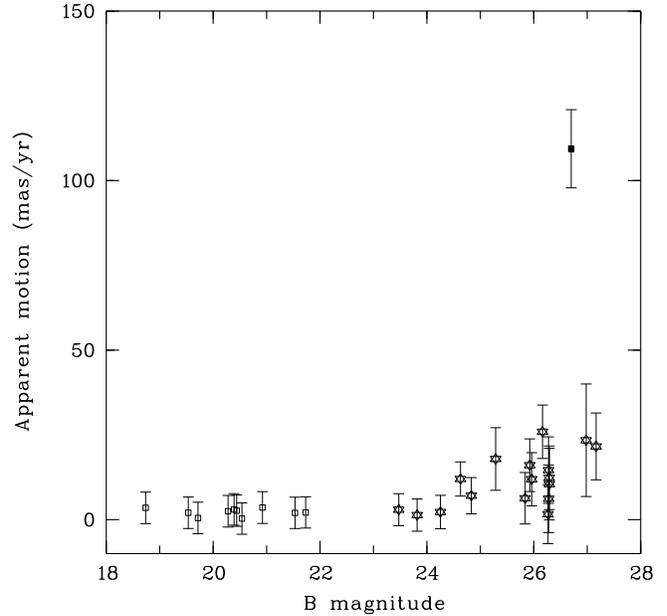,angle=-90,bbllx=10pt,bblly=60pt,bburx=590pt,bbury=620pt,width=8.8cm,clip=true}

\caption[]{Apparent proper motion of local astrometric reference stars (open squares),
comparison objects (stars) and of \rx \ (filled square) using December 2000 and December
2002 - January 2003 VLT images.}

\label{Bpm}
\end{figure}

We searched for possible proper motion by comparing the positions of the optical counterpart at
five epochs from 1997 till 2003. Apart from the 1997 images which are the sum of data collected
over a time interval of 2 months from February 7 till April 4, the four other epoch observations 
took  place during integer number of consecutive years within 27 days from the same calendar date.
Overall, the change in position due to trigonometric parallax is expected to be of the order of 16
mas at maximum for an assumed distance of 100\,pc (mainly due to the spread in time of the 1997
observations), and can therefore be neglected for our purpose. As for the photometry we found that
fitting 2-D Gaussian profiles was providing the most accurate results and good estimates of the
positional errors. Positions from automated processes such as Sextractor or DAOPHOT turned out to
be less reliable in general. The fixed width of the stellar profiles in right ascension and
declination were first derived by averaging the values obtained on a set of relatively bright but
nonsaturated stars. Then the position of a group of 10 reference stars was measured, together with
that of \rx \ and of 17 faint test objects. Reference stars are evenly spread within 30\arcsec \ of
\rx \ and are in the range of B magnitude from 21.8 to 18.8. Comparison objects are distributed in
the same sky region as the reference stars and were selected for their stellar-like appearance.
Their magnitudes range from B $\sim$ 27.0 to 23.5. Using a rotation, translation and scaling
transformation, the 1997, 1999, 2000 January and 2003 positions were then moved to the 2000
December FORS1/VLT reference frame which provides the most accurate positions. Not all the
reference stars are common to all frames and the 1997 to December 2000 transformation was computed
with only 8 stars. We checked that the quality of the transformation was not depending heavily on
the number of astrometric reference stars used. Since part of the proper motion could be due to
bulk motion of the reference stars we also computed the astrometric solution of the December 2000
VLT image using 78 USNO-A2 field stars. Reference stars do not show any systematic difference in
position in the time interval from 1980 (USNO-A2) till 2000.  The accuracy with which reference
stars coincide after transformation is rather good. Their maximum offset to mean positions after
transformation are listed in Table \ref{astroerrors}. Between December 2000 and 2002-2003 VLT
observations, the mean residual of the 10 reference stars is only 4.5 mas. We plot in
Fig.~\ref{Bpm} the apparent displacements of the 17 faint comparison stars together with those of
\rx \ and of the astrometric references derived from the two FORS1/VLT observations.  None of the
reference stars exhibit significant motion with respect to the rest of the group. Compared to stars
of similar brightness, the apparent motion of \rx \ is clearly of high significance. 

\begin{table}
\caption[]{Maximum errors (milli-arcseconds) in RA and Dec for the astrometric reference stars}
\label{astroerrors}
\begin{tabular}{ccc}
\noalign{\smallskip}
Epoch           &   RA            & DEC\\ 
\hline
Feb-Apr  1997   &  12.0    &  4.2 \\
January  1999   &  4.4     &  4.3 \\       
January  2000   &  32      &  34  \\ 
December 2000   &  4.1     & 7.0  \\  
Dec 2002 Jan 2003   &  5.2     & 3.2  \\         
\noalign{\smallskip}
\hline
\end{tabular}
\end{table}

We also corrected for differential refraction effects resulting from the large colour difference
between \rx \ (U-B = -1.11) and the reference astrometric stars (mean U-B = 0.17). In the B filter, this
translates into a shift of $\sim$ 130\AA \ in effective wavelength. Since all observations were done
at low airmass, the corrections remain small, of the order of 5 to 30 mas and are usually negligible
compared to positioning errors. Differential refraction effects are only significant in the case of the
December 2000 VLT observation.

\begin{figure}
\psfig{figure=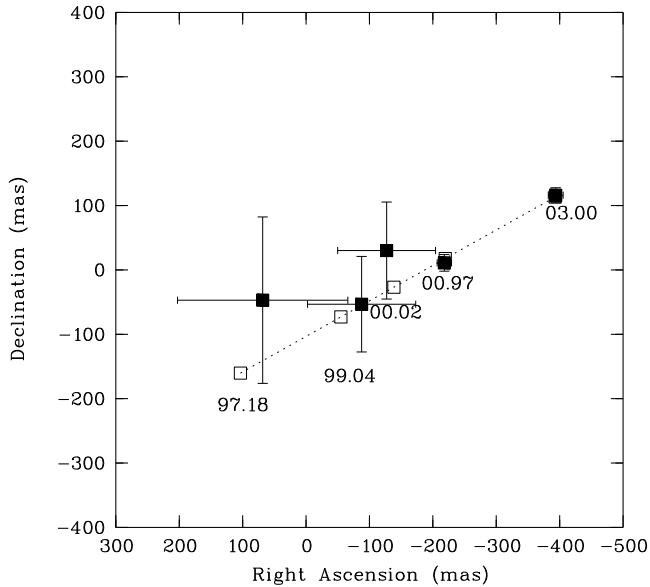,angle=-90,bbllx=0pt,bblly=30pt,bburx=590pt,bbury=620pt,width=8.8cm,clip=true}
\caption[]{Positions of \rx \ for the five epochs (filled squares) together with the best fit proper motion
solution (open squares) corresponding to $\mu$ = 97 mas/yr}
\label{pm}
\end{figure}

We then fitted a proper motion solution expressed in terms of total motion and direction angle to the
various positions of \rx \ using a minimum $\chi^{2}$ algorithm. The best fit shown in Fig. \ref{pm} is
obtained for $\mu = 97\pm 12$ mas/yr. The direction of motion is oriented  29\degree\ $\pm$
7\degree\ north of west (all errors are 90\% confidence level for one parameter).

\section{Optical to X-ray energy distribution}

In the X-ray domain \rx\ was observed with the ROSAT, SAX, XMM-Newton and Chandra observatories. 
As results inferred from different instruments are subject to calibration uncertainties (see,
e.~g., Burwitz et al. \cite{burwitz03} for the case of \rxa), we used only the Chandra data
collected with the   LETGS grating instrument in a total exposure of 37.5 ks in February 2000.
First results from the spectral analysis of these data have been presented in Pavlov et al.
(\cite{pavlov2002}). They found that neither light-element (hydrogen or helium) nor heavy-element
(pure iron or solar mixture) neutron star atmosphere models can explain the multi-wavelength
spectral distribution of the emission detected from \rx.  The net result of this analysis was
remarkably similar to the situation with the spectral fitting of the optical and X-ray data on
\rxa: while providing acceptable fits to the X-ray data the light-element atmosphere models  yield
unrealistic large radii of the INS and significantly over-predict the actual optical fluxes (by a
factor of $\sim 300$), whereas the heavy-element models do not fit the X-ray data due to numerous
features in the model spectra caused by absorption at atomic transitions in ions of the heavy
elements (see Zavlin \& Pavlov \cite{zavlin02} for details). Pavlov et al. (\cite{pavlov2002})
found that the best fit to the Chandra LETGS data on \rx\ is provided by a blackbody model with the
temperature kT$^\infty_{\rm bb}=80-82$ eV emitted from an area of  a radius R$^\infty_{\rm
bb}=(2.0-2.2)\,({\rm D}/100\,{\rm pc})$ km (D is the distance to the source), absorbed with the
hydrogen column density  N$_{\rm H}=(1.5-1.7)\times 10^{20}$ cm$^{-2}$.  

Similar to the case of \rxa, the best blackbody model derived from the X-ray data on \rx\ underpredicts
the optical fluxes detected from the source by a factor of 4. Pavlov et al. (\cite{pavlov2002}) 
speculated that a two-blackbody model, originally suggested by  Pons et al. (\cite{pons02}) to describe
the multi-wavelength data on \rxa, may be also applied to the radiation from \rx, with the ``soft''
component of kT$^\infty_{\rm bb,s}<43$ eV and R$^\infty_{\rm bb,s}>6.1\,({\rm D}/100\,{\rm pc})$ km.  Such
a model, with X-rays originating from a hot area on the star's surface and the optical fluxes emitted from
the rest of the surface, may explain the pulsations of the X-ray radiation from \rx. In addition to
the thermal components,  HST observations by Kaplan et al. (\cite{kaplan03}) reveal evidence for a
nonthermal power law component.

However, while this simple multiple-component model seems to be in agreement with the
properties of the emission detected from \rx, it is hardly reconciled with the fact that, as stars
are not blackbodies, radiation emitted by a star should deviate from a blackbody model. Also, in the
case of \rxa , the absence of pulsations put severe constraints on the two-blackbody model, requiring
either a particular geometrical configuration or strong gravitational deflection (Ransom et al.
\cite{ransom02}). 

All previous  models of neutron star surface radiation were based on a conventional assumption that
there is enough matter (e.~g., pure hydrogen or iron, or a mixture of elements)  on the neutron
star surface to  make the atmosphere layers optically thick at all energies of interest. A
typical estimate for such an amount in terms of the total surface column density is y$_{\rm
tot}>10-100$ g cm$^{-2}$  (depending on the surface temperature) to provide the equilibrium 
(or diffusion)  solution of the radiative transfer problem in the very deep layers (see Mihalas
\cite{mihalas78}). Under this assumption, spectra of the emitted radiation are solely
determined by the temperature distribution in the atmosphere (which grows towards larger depths)
and do not depend on properties of star's layers lying underneath the atmosphere. In the case of a
light-element atmosphere composition, the model spectra were found to be much harder at higher
photon energies than blackbody ones providing the same radiative flux. The reason for this effect
is that high-energy photons with longer mean-free-paths are emitted from deep surface layers
with temperatures larger than the so-called effective temperature, the fourth power of which
determines the total energy flux (see, e.g., Zavlin et al. \cite{zavlin96} and  Zavlin \& Pavlov
\cite{zavlin02} for details). Due to this property  the model spectra, when applied to
observational data, usually yield lower temperatures by a factor of $2-3$ and much larger emitting
areas by a factor of $50-200$  than estimates obtained from blackbody fits. However, it is not
known a priori how much, for example, hydrogen has been deposited on  the surface of a neutron star
(e.~g., by accretion from the interstellar medium or from a fallback disk). If the hydrogen
layer is not optically thick at all energies, but only at lower ones, then the atmosphere structure
(temperature run) is modified, that in turn should affect the spectra of emergent radiation.

\begin{figure}

\psfig{figure=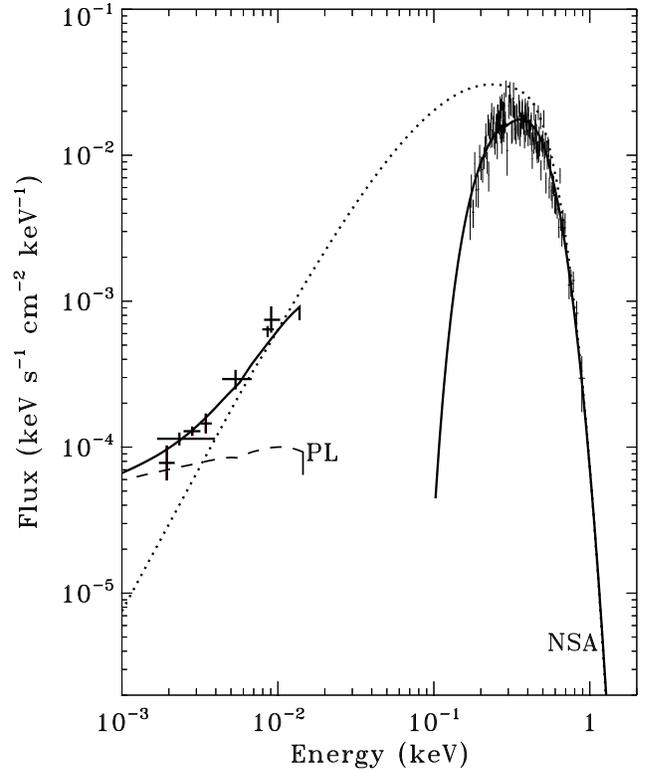,height=11cm}

\caption[]{ Multiwavelength  energy distribution of \rx. The optical and far UV data are from Kulkarni
\&  van Kerkwijk (1998), Kaplan et al. (2003) and this paper. The X-ray spectrum is that from the
Chandra LEGTS data. Solid line shows the best thin neutron star atmosphere (NSA) + nonthermal power
law (Kaplan et al. 2003, dashed line) model fitting both the X-ray and optical data (see text for the
parameters). The dotted line gives unabsorbed fluxes.}

\label{rxj0720_spec}
\end{figure}

We investigated nonmagnetic atmosphere models with various values of the total column density of
hydrogen y$_{\rm tot}$ in the atmosphere. We assumed the atmosphere to be in the radiative and
hydrostatic equilibrium, and that the radiation at the inner boundary y=y$_{\rm tot}$ is given by the
diffusion solution (although, the latter condition is not strictly justified --- see Discussion).
The main result we obtain is rather natural: reducing the total column density lowers the temperature
of surface layers which emit high-energy photons, that in turn leads to the softening of the Wien
tail in the model spectra, making them close to the blackbody ones with the same effective
temperature. This means that optical fluxes predicted by these models applied to X-ray data on INSs
are expected to be significantly decreased in comparison with those given by the standard model, but
still larger than values yielded by the blackbody model.  We computed hydrogen models in a grid of
effective temperature and  total column density (down to y$_{\rm tot}=10^{-3}$ g cm$^{-2}$) and
applied these models to the UV - optical and Chandra LETGS data on \rx \ taking into account the
nonthermal power law discovered by Kaplan et al. (\cite{kaplan03}). A best fitting model is shown
in Fig.~\ref{rxj0720_spec}. The parameters of the model are: total atmosphere column density y$_{\rm
tot}=0.16$ g cm$^{-2}$, effective temperature kT$_{\rm eff}=57$ eV, distance to the star  D~=~204 pc
(assuming the standard neutron star radius R~=~10~km and mass M~=~1.4$\,{\rm M}_\odot$),  and
interstellar absorption N$_{\rm H}=1.1\times 10^{20}$ cm$^{-2}$.  As seen in Fig.~
\ref{rxj0720_spec}, the model nicely fits  the LETGS data ($\chi^2_\nu$=0.94 in the X-ray domain) and
agrees well with the detected optical fluxes. The total mass of hydrogen obtained in the fit is
4 $\pi$R$^{2}$ y$_{\rm tot}$ = 2$\times$10$^{12}$ g. Assuming a stellar age of 10$^{6}$ yr, this
translates into a mean accretion rate of $\sim$ 0.6\,g/s, which is many orders of magnitude
lower than predicted by the accretion model.

\section{Discussion}

Despite the fact that the thin atmosphere model provides a good fit to the observational data in the
broad energy range at reasonable neutron star parameters, it serves rather as an illustrative example
because of a number of shortcomings in the simplified approach it has been derived with. First of
all, the assumption on the equilibrium solution at the maximal depth in such an atmosphere may not be
justified at higher X-ray energies where the emitted radiation could be affected by properties
of the stellar surface under the atmospheric layer. It means that the boundary condition at  the
inner boundary y$={\rm y}_{\rm tot}$, taken currently to be the same as in the standard modeling 
(with much larger values of y$_{\rm tot}$), may need to be modified. For instance, one could consider
such a thin atmosphere on top of a solid neutron star surface or a thicker layer composed of another
chemical element. Second, the model should be developed with account for the presence of a strong
surface magnetic field, $B\sim 10^{13}$ G, as recently suggested by Zane et al. (\cite{zane02}).
It is safe to assume that in the case of magnetized models, reducing the atmosphere thickness would
also soften the spectrum of the emergent radiation. Since in strong magnetic fields the emergent
spectra of the standard models are intrinsically softer than the non-magnetic ones (Zavlin \& Pavlov
2002), one may expect that the magnetized model needs a larger atmosphere thickness in order to
describe the broad-band properties of the emission from \rx .

Another point here is that this model is constructed assuming a uniform
neutron star surface, which makes the model emission unpulsed, whereas the observed X-rays from
\rx\ reveal pulsations with a pulsed fraction of about 10\%. A way to model the pulsations is to
consider a nonuniform temperature distribution over the surface (see, e.g., Cropper  et al.
\cite{cropper01}) since such a temperature nonuniformity  can be induced by the strong magnetic
field  (Greenstein \& Hartke \cite{green83}). Finally, in its present state, the thin atmosphere
model applied to the multiwavelength data on \rxa \ underpredicts the distance  to this object by a
factor of 2. In light of all this, the presented spectrum should be considered as a first, perhaps
rather crude step towards more advanced modeling of radiation from a neutron star surface covered
with a thin atmosphere layer, which might help to elucidate the origin of thermal emission detected
from INSs.

A proper motion of 97 $\pm$ 12 mas per year corresponds to a transverse velocity of V$_{\rm T}$
$\sim$ 50 $\times$ (D/100\,pc) km\,s$^{-1}$. A small part of this displacement could be due to solar
motion (about 12 mas/yr in the right ascension if the source is at a distance of 100\,pc). \rx \ is
thus the second radio-quiet soft X-ray bright isolated neutron star for which proper motion is
detected. The value reported here is about a third of that measured for \rxa \ 
(Neuh\"auser~\cite{neu01}; Walter \cite{walter2001}; Kaplan et al. \cite{kaplan02b}).

Such a spatial motion, although modest for neutron star standards, most probably rules out the
accretion scenario for \rx . The observed bolometric luminosity of $2.6\times 10^{31}$
(D/100\,pc)$^{2}$ \ergs\ (Haberl et al. \cite{haberl97}) implies accretion rates of the order of
$\dot{\rm M}$ $\sim$ 1.2 10$^{11}$ (D/100\,pc)$^{2}$ g\,s$^{-1}$.  On the other hand, the accretion
rate can be estimated   assuming the Bondi-Hoyle model:  $\dot{\rm M}$ $\sim$ $n$ v$_{10}^{-3}$
10$^{11}$ g\,s$^{-1}$ where $n$ is the mean density of the interstellar medium  (in cm$^{-3}$) and
v$_{10}$ the relative velocity in units of 10 km\,s$^{-1}$ (Treves et al. \cite{treves2000}).
Therefore, very high ISM densities of the order of $n \ \sim\ $150 (D/100\,pc) cm$^{-3}$ are required
to account for the observed X-ray flux in the accretion scenario. This density is much higher than the
mean density derived from  soft X-ray photoelectric absorption or from optical reddening of field
stars, $n$ $\sim$ 0.2 - 0.4 cm$^{-3}$ (Motch \& Haberl \cite{mh98}). However, there have been recent
reports on evidence for dense ISM structures ($n$ $\sim$ 10 - 1000 cm$^{-3}$) on small scales ranging
from 10 to 10$^{6}$ AU in regions of otherwise low mean densities (Lauroesch \& Meyer \cite{lmeyer} and
Meyer \& Lauroesch \cite{meyerl}). \rx \ could be presently passing through one of these high density
structures at a velocity of $\geq$ 10 $\times$ (D/100\,pc) AU yr$^{-1}$. Since such high density clouds
are likely to be rare, we would only detect X-ray emission from old neutron stars when they pass
through these small regions. The X-ray ionization of these cloud-lets may be detectable in the optical.
It is not clear, however, whether the apparent absence of X-ray flux and temperature changes from \rx \
is compatible with the size of these dense structures and the expected time scale for accretion
instability. This scenario furthermore, cannot account for the large $\dot{\rm P}$ $\geq$
10$^{-14}$\,s\,s$^{-1}$ measured by Zane et al. (\cite{zane02}).

Another implication of the high proper motion is that in no case can the spin-down derived by Zane et
al. (\cite{zane02}) be due to a propeller torque extracted from the interstellar medium. Following
Colpi et al. (\cite{colpi98}), propeller spin-down can be written as $\dot{\rm P} \ \sim \ 3.1\times
10^{-16}\, n^{9/13} \, {\rm v}_{10}^{-27/13}\, B_{12}^{8/13} \, {\rm P}^{21/13} \, \rm s\, s^{-1}$
where  $B_{12}$ is the polar magnetic field in units of 10$^{12}$ G.  At $\dot{\rm P}$ = 5.4
10$^{-14}$\,s\,s$^{-1}$ the propeller mechanism dominates over the magnetic dipole losses if v$_{10} \
\leq \ 0.9\, n^{1/3}$. Since v$_{10}$ is likely greater than unity, the propeller mechanism could only
dominate if the star presently goes through a high density medium with $n$ $\geq$ 100.

The possibility that \rx \ breaks through propeller interaction with a fallback disk seems as
well unlikely. First, assuming that the entire X-ray luminosity is due to friction leads to much
larger $\dot{\rm P}$ than observed (Zane et al. \cite{zane02}; Kaplan et al. \cite{kaplan02}).

In the framework of the blackbody interpretation of the X-ray data, a part of the optical flux excess
above the extrapolation of this model to the low-energy band (see \S~5 and Fig.~\ref{rxj0720_spec})
could arise from the reprocessing of X-rays in a residual debris disk when the neutron star has
entered the pulsar phase. Emission spectra of such discs were computed by Perna et al.
(\cite{perna00}) in a number of configurations. Although in the B band, the entire flux excess above
the X-ray blackbody extrapolation could be due to a remnant disc, its energy distribution is rather
red and would fail to account for the U band excess. Therefore, it is not clear whether such discs
can significantly contribute to the overall optical emission of \rx . Similar conclusions are
reached by Kaplan et al. (\cite{kaplan03}).

The lack of parallax and radial velocity as well as the large uncertainty on the parameters of the proper
motion do not allow to guess the birth place of \rx \ as it is apparently possible for \rxa\ (Walter \&
Lattimer \cite{wl2002}). The star is moving away from the Galactic plane but is still very close to it.
Its general direction of motion is consistent with a birth either in the Sco OB2 complex, as for \rxa\
(more probably in the lower Centaurus Crux region at D $\sim 170$ pc), or in the Vela OB2 + Trumpler 10
association (D$=360-410$ pc). In both cases, the flight time from birth place is of the order of
10$^{6}$~yr with a rather large uncertainty. This value is generally consistent with the spin down age
and with cooling time (see Zane et al. \cite{zane02}).

Several high magnetic field (B $\geq$ 10$^{13}$\,G) radio pulsars with long periods have been recently
discovered (Camilo et al. \cite{camilo00}; Morris et al. \cite{morris02}). One of these radio pulsars,
PSR J1830--1135 has properties astonishingly close to those of \rx \ (P = 6.22\,s,   $\dot{\rm
P}=4.7\times 10^{-14}$\,s\,s$^{-1}$  and $B=1.7\times 10^{13}$ \,G, Morris et al. \cite{morris02}). It
could thus well be that \rx \ and several other X-ray bright and radio-quiet INSs are radio pulsars whose
radio beams do not cross the Earth (Motch \cite{mo2001}; Kaplan et al. \cite{kaplan02}, Kaplan et al.
\cite{kaplan03} ). The effect of radio beam narrowing with increasing  periods (Biggs \cite{biggs90})
could plausibly explain the much smaller ratio of long-period to short-period radio pulsars compared to
that found for X-ray pulsars.

\section{Conclusions}

ESO-VLT and 4m class telescope imaging reveal that \rx \ has a proper motion of $\mu$ =  $97\pm 12$
mas/yr. Such a relatively high spatial velocity almost completely rules out the possibility
that the neutron star accretes matter from the interstellar medium. This strongly favours the
idea that \rx \ is a cooling INS of an age around 10$^6$ yr. A propeller mechanism on a fallback
disk, or on accreted material from the ISM cannot account for the braking detected by Zane et al.
(\cite{zane02}) and thus the observed spin-down is likely due to magnetic dipole losses.  Recent
discoveries of radio pulsars with comparable periods and magnetic fields combined with the effect of
the radio beam narrowing at longer periods suggest that \rx \ could well be an off-beam radio pulsar.
As in \rxa, the optical continuum lies well above the extrapolation of the blackbody model seen in
X-rays.  To account for the optical emission  one can add a lower temperature blackbody component
cold enough not to be detected in the X-ray range (Pavlov et al. \cite{pavlov2002}, Kaplan et
al. \cite{kaplan03}). However, neutron stars are unlikely to radiate like blackbodies. Spectra
predicted by  heavy-element  atmosphere models have  a number of prominent features  which should be
conspicuous in  the Chandra and XMM-Newton data obtained with high-energy resolution. On the other
hand, the standard (thick) hydrogen atmosphere models overpredict the optical emission by several
magnitudes. We propose here a new kind of model atmosphere in which the thickness of the
light-element layer has reduced values. Although the model still has a number of shortcomings,  it
can adequately fit the overall energy distribution with a single component,  requiring the distance
to \rx\ to be about 200 pc which still has to be compared with the actual value yielded by future
parallax measurements.

%%%

\begin{acknowledgements}
We thank the referee, Patrick Slane for very valuable comments. 
VEZ acknowledges discussions on
the atmosphere model modifications with Bernd Aschenbach,
George Pavlov and Joachim Truemper.
\end{acknowledgements}

\end{document}